\documentclass[12pt]{article}

\usepackage{epsfig}
\usepackage{cite}
\usepackage{amsmath, amssymb, amsfonts}
\usepackage{color}
\usepackage{latexsym}
\usepackage{graphicx}
\usepackage{cancel}
\usepackage[colorlinks,bookmarks]{hyperref}
\hypersetup{pdfpagemode=UseNone, pdfstartview=FitH, linkcolor=blue,
            citecolor=red, urlcolor=blue}

\def\be{\begin{equation}}
\def\ee{\end{equation}}
\def\ba{\begin{eqnarray}}
\def\ea{\end{eqnarray}}

\def\bdm{\begin{displaymath}}
\def\edm{\end{displaymath}}
\def\la{~\mbox{\raisebox{-.6ex}{$\stackrel{<}{\sim}$}}~}
\def\ga{~\mbox{\raisebox{-.6ex}{$\stackrel{>}{\sim}$}}~}
\def\bq{\begin{quote}}
\def\eq{\end{quote}}

 at 10truept

\renewcommand{\(}{\left(}
\renewcommand{\)}{\right)}

\newcommand{\Mpl}{M_{\mathrm{Pl}}}

\newcommand{\bea}{\begin{eqnarray}}
\newcommand{\eea}{\end{eqnarray}}

\newcommand{\bi}{\begin{itemize}}
\newcommand{\ei}{\end{itemize}}

\newcommand{\beq}{\begin{equation}}
\newcommand{\eeq}{\end{equation}}
\newcommand{\beqa}{\begin{eqnarray}}
\newcommand{\eeqa}{\end{eqnarray}}

\def\la{~\mbox{\raisebox{-.6ex}{$\stackrel{<}{\sim}$}}~}
\def\ga{~\mbox{\raisebox{-.6ex}{$\stackrel{>}{\sim}$}}~}

\def\12{{1 \over 2}}

\def\ltap{\ \raise.3ex\hbox{$<$\kern-.75em\lower1ex\hbox{$\sim$}}\ }
\def\gtap{\ \raise.3ex\hbox{$>$\kern-.75em\lower1ex\hbox{$\sim$}}\ }
\def\gl{\ \raise.5ex\hbox{$>$}\kern-.8em\lower.5ex\hbox{$<$}\ }
\def\roughly#1{\raise.3ex\hbox{$#1$\kern-.75em\lower1ex\hbox{$\sim$}}}

\begin{document}

\thispagestyle{empty}
\begin{flushright}
July 2022 
\end{flushright}
\vspace*{1cm}
\begin{center}
  
{\Large \bf Power-law Inflation Satisfies Penrose's Weyl}
\vskip.2cm
{\Large \bf  Curvature Hypothesis}

\vspace*{.55cm} {\large Guido D'Amico$^{a, }$\footnote{\tt
damico.guido@gmail.com} and Nemanja Kaloper$^{b, }$\footnote{\tt
kaloper@physics.ucdavis.edu}
}\\
\vspace{.5cm} {\em $^a$Dipartimento di SMFI dell' Università di Parma and INFN}\\
\vspace{.05cm}{\em Gruppo Collegato di Parma, Italy}\\
\vspace{.3cm}
{\em $^b$QMAP, Department of Physics and Astronomy, University of
California}\\
\vspace{.05cm}
{\em Davis, CA 95616, USA}\\

\vspace{1cm} ABSTRACT
\end{center}
Based on entropy considerations and the arrow of time Penrose argued that the
universe must have started in a special initial singularity with 
vanishing Weyl curvature. This is often interpreted to be at odds with 
inflation. Here we argue just the opposite, that Penrose's persuasions are in 
fact consistent with inflation. Using the example of power law 
inflation, we show that inflation begins with a past null singularity,
where Weyl tensor vanishes when the metric is initially exactly conformally
flat. This initial state precisely obeys Penrose's conditions. 
The initial null singularity breaks 
$T$-reversal spontaneously and picks the arrow of time. It can be regulated and interpreted as a
creation of a universe from nothing, initially fitting in a bubble of Planckian size when it 
materializes. Penrose's initial conditions are favored by the initial $O(4)$ symmetry 
of the bubble, selected by extremality of the regulated Euclidean action. The predicted observables are 
marginally in tension with the data, but they can fit if small corrections to power law inflation 
kick in during the last 60 efolds.

\vfill \setcounter{page}{0} \setcounter{footnote}{0}

\vspace{1cm}
\newpage

\section{Introduction}

Complex things which break don't reassemble on their own. Putting them together takes a toll
and this toll is exacted by the increase of entropy of the system describing the process
of breaking and reassembly. This trend can be used to define a global arrow of time
in the universe, in contrast to generic microphysical phenomena which typically respect 
time reversal. 

Penrose has taken this observation a step further \cite{Penrose:1979azm}, arguing that this phenomenon 
implies that the universe originated 
from a special initial state,  characterized by a singularity in whose vicinity the
geometry of the universe is very well approximated by conformal flatness, with (almost)
vanishing Weyl tensor. This is also often interpreted as a problem for inflation (for a range of viewpoints, see 
\cite{Penrose:1988mg,Davies:1983nf,Page:1983uh,Davies:1984qc,Albrecht:2002uz,Albrecht:2004ke,Carroll:2004pn,Carroll:2005it,Gibbons:2006pa}). Recall that 
the idea of inflation \cite{Guth:1980zm,Linde:1981mu,Albrecht:1982wi}
is to blow up a universe from an initially small region, whose initial contents is much
smaller than the vast complexity observed in the universe today. This is regardless of how the contents is inventoried,
naively by counting over the initial volume, or more consistently by using the initial apparent horizon size. Either
way, the late universe has far more contents than the early one. The difficulty with this obvious fact 
is that something other than inflation seems to be needed to select this seemingly improbable 
initial state. In other words, if the entropy count is used as a measure of likelihood, it 
seems to suggest that inflation presupposes an
unlikely initial state. 

Curiously, this argument overlooks the simple experiential fact that in many models of inflation
the initial state of inflation is both singular and has an almost conformally flat geometry, in full accord with the
technical aspects of Penrose's hypothesis. Indeed, the now-classic Borde-Guth-Vilenkin theorem asserts that inflationary
spacetimes are past geodesically incomplete \cite{Borde:2001nh} (see also \cite{Lesnefsky:2022fen}), which at least at the semiclassical gravity level implies
that inflation starts out of a singularity. Moreover, once inflation sets in\footnote{A careful critic would without doubt
express a concern right now that maybe inflation never sets in. We postpone our reply aimed at dispelling this concern for
later in this paper.}, it quickly dilutes initial deviations from 
homogeneous and isotropic FRW metric 
\cite{Wald:1983ky,bartip1,bartip2,Kaloper:2002uj,Kaloper:2002cs,East:2015ggf,Kleban:2016sqm,Clough:2016ymm,Kaloper:2018zgi,Burgess:2020nec}, 
which being conformally flat has vanishing Weyl tensor, by symmetry. 
Thus it seems that at least `mechanically', if we accept Penrose's argument
that the initial state is singular and Weyl flat, it is completely consistent to get inflation to 
spring forth from it. In some sense, actually, this state would appear to favor subsequent inflation
as the origin of observed structures, since Weyl flatness favors a very smooth initial universe
and something is required to break that smoothness spontaneously, instead of explicitly -- precisely what inflation is intended to do. 

To make our point, we employ the example of power law 
inflation \cite{Mathiazhagan:1984vi,Lucchin:1984yf,Liddle:1988tb,Kalara:1990ar}. 
We explain that the inflationary past ultimately 
begins with a past null singularity \cite{Hellerman:2001yi,Fischler:2001yj}, for both spatially flat and spatially open FRW
metrics. Since both of these metrics are initially exactly conformally flat, they have vanishing Weyl tensor. Clearly, 
the initial null singularity breaks time-reversal and picks the arrow of time. Thus both of these metrics, maximally extended
into the past satisfy Penrose's Weyl curvature hypothesis and hence describe universes with an arrow of time which 
nevertheless inflate. We don't think our examples are unique. Other examples may also exist, which feature spacelike instead of null singularities, such
as the closed universe arising from the instantons in the no-boundary proposal. In fact, when we regulate 
the null singularity examples, which we consider in detail below, the regulators {may be} spacelike surfaces, 
which we comment later on. 
The point we are trying to make, however, is that regardless of the specific
nature of the singularity, the selection of the initial state which realizes Penrose's Weyl curvature hypothesis might
be a consequence of the quantum completion of inflation, which is anyway necessary, instead of needing a 
completely separate mechanism. 

The question about what specifically selects the initial singularity can be addressed using quantum cosmology 
and no-boundary proposal. The past null singularity can be understood in terms of  
the singular Hawking-Turok instantons \cite{Hawking:1998bn,Turok:1998he,Linde:1998gs,Unruh:1998wc,Vilenkin:1998pp}\footnote{A different method to start the universe with a null singularity has been proposed 
in \cite{Aguirre:2003ck}.}, 
which can be regulated and interpreted as an expanding nonsingular 
bubble (for various approaches see \cite{Garriga:1998tm,Garriga:1998ri,Bousso:1998pk,Blanco-Pillado:2011fcm,Brown:2011gt}). Using this approach 
gives the reason for the selection of the initial Weyl-flat state of inflation: it minimizes the
Euclidean action thanks to the $O(4)$ symmetry of the configuration and the smallness of the primordial bubble which seeded the
universe \cite{Coleman:1980aw,Linde:2004nz}. 

The model actually yields predictions close to the current
BICEP/Keck bounds \cite{BICEP:2021xfz}, which can be improved with 
small corrections\footnote{{We will ignore the specific form of those corrections here, and work with purely exponential 
potentials because the causal structure analysis is considerably simpler.}}
to the potential during the last 60 efolds. Alternatively,
if the resolution of the $H_0$ tension is Early Dark Energy (EDE) 
\cite{Poulin:2018cxd,Niedermann:2019olb,Niedermann:2020dwg}, the CMB fits need a 
slightly higher scalar spectral
index $n_{\tt S} \sim 0.98$ -- $0.995$ \cite{Ye:2021nej,Takahashi:2021bti,DAmico:2021fhz}, 
which is readily retrofitted by power law inflation. Interestingly, for the parameters which are close
to the observationally favored values, the regime of universe self-reproduction in power law inflation is relegated to the cutoff 
physics, and so are superseded by the primordial bubble. This means, once fixed by the birth of the universe, the arrow of time
remains unaffected by subsequent dynamics. 

A very interesting question is how to interpret the cosmological perturbations, both scalar and tensor,
which arise during inflation from the entropic point of view. Scalar perturbations are model dependent, although
in all models of inflation they are an intrinsic ingredient of inflationary dynamics. Tensor perturbations are
however universal, depending only on the scale of inflation. Both modes however utilize the same ``seed", which 
is the uncertainty principle of quantum fluctuations in the inflationary vacuum. In (quasi)-de Sitter geometries
this leads to the spontaneous emergence and growth of anisotropies and inhomogeneities, which may be
viewed as an avatar of de Sitter instability \cite{Mukhanov:1996ak,Polyakov:2007mm,Dvali:2017eba,Kaloper:2022oqv,Kaloper:2022utc,Kaloper:2022jpv}. This instability, from the entropic 
point of view, indicates that the pure de Sitter, appearing as the state with vanishing Weyl curvature, is a special
state of the theory that dynamically evolves into the more generic states, which include the perturbations. 
It would be interesting to test this idea in more detail.

\section{Power Law Inflation}

Power law inflation is driven by a scalar field with an exponential potential, 
with the field rolling off to $\phi = \infty$. The potential is parameterized 
by \cite{Mathiazhagan:1984vi,Lucchin:1984yf,Liddle:1988tb,Kalara:1990ar} 
\be
V(\phi) = V_0 e^{-  c \phi/\Mpl} \, ,
\label{qpot}
\ee
where $\phi$ is a canonically normalized scalar field, 
$c$ is a numerical constant of order unity, and $\Mpl \sim 2 \times 10^{18} ~ {\rm GeV}$ is the
Planck scale. Clearly, $V_0$ is degenerate with the initial value of $\phi$. 
Alternatively, the dynamics 
can also be parameterized by an equation of state 
\be
p= w \rho \, ,
\label{eqstate}
\ee
where $p$ and $\rho$ are pressure and energy density, respectively. 
In general, for homogeneous solutions the equation of state parameter
$w$ is a function of time until the self-similar attractor is reached. When 
$c\ll \sqrt{2}$ (which is the requirement that the geometry describes an
accelerating expansion) a typical configuration will settle into the attractor
fixed point within a few Hubble times, and $w \rightarrow {\rm const.}$ 

The scalar sources the FRW metric
\be
ds^2 = -dt^2 + a^2(t)\Bigl( \frac{dr^2}{1-kr^2} + r^2 d\Omega_2 \Bigr) \, .
\label{met}
\ee
Here we will be particularly interested in the $k = 0, -1$ cases, with spatially flat or open hyperbolic slices.
Equations of motion are 
\be
3H^2 + 3\frac{k}{a^2} = \frac{\rho}{\Mpl^2} \, , ~~~~~~~  \dot \rho + 3H(\rho + P) = 0 \, , 
~~~~{\rm with}~~~~ \rho = \frac{\dot \phi^2}{2} + V \, , ~~~~~~~ p = \frac{\dot \phi^2}{2} - V \, ,
\label{eoms}
\ee
where the Hubble parameter is $H = \dot a/a$. 

To find the attractor, we substitute $\rho$ and $p$ into (\ref{eqstate}) and hold $w$ fixed,
which gives the first order equation $\dot \phi^2/2 = \frac{1+w}{1-w} V$. This is easy to solve; after
straightforward algebra, we find the attractor form of $\rho$ (with $p= w \rho)$,  
\be
\rho = \frac{4}{c^2} \frac{1}{1+w} \frac{\Mpl^2}{t^2}  \, .
 \label{attractor} 
\ee
Next, the conservation equation yields $\rho = \rho_0 (a_0/a)^{3(1+w)}$, and so comparing with (\ref{attractor}) we find
$a \sim t^{\frac{2}{3(1+w)}}$. The Friedmann equation then shows that unless $w=-1/3$, the curvature contribution is subleading
relative to the attractor energy density. Neglecting it and substituting $\rho$ of (\ref{attractor}) into it yields, using $H = \frac{2}{3(1+w)t}$,  
\be
1+w = \frac{c^2}{3} \, .
\label{wc}
\ee
Clearly, imposing $w \rightarrow - 1$ requires $|c|\ll 1$. In any case, the attractor is 
\cite{Mathiazhagan:1984vi,Lucchin:1984yf,Liddle:1988tb,Kalara:1990ar} (since $\frac{2}{3(1+w)} = \frac{2}{c^2}$,
and using $\frac{\dot \phi^2}{2} = \frac{2}{c^2} \frac{\Mpl^2}{t^2}$)
\be
a = a_0 \bigl(\frac{t}{t_0}\bigr)^{\frac{2}{c^2}} \, ,  ~~~~~~~~~~ \phi = \phi_0 + \frac{2 \Mpl}{c} \ln(\frac{t}{t_0}) \, .
\label{attsoln}
\ee
Here $a_0, t_0$ and $\phi_0$ are integration constants; $a_0$ is pure gauge, which we can fix to unity choosing $a(t_0) = 1$. The
others satisfy $V_0 e^{-c \phi_0/\Mpl} t_0^2 = 2 \Mpl^2(6-c^2)/c^4$. 

This solution applies at late times. At early times, it may be altered at small $t$ If the universe is spatially curved, specifically open, with $k=-1$,
and the curvature initially dominates. In that case, the scale factor changes to $a = t/t_0$, while the scalar field configuration remains 
largely the same.  

In either case, it is evident that $t \rightarrow 0$ is an initial singularity. In fact, when $w \le -1/3$, 
this hypersurface is null \cite{Hellerman:2001yi,Fischler:2001yj}, as we will review below. Here we merely
note that the requirement of using Einstein's equations consistently near the singularity imposes a physical
cutoff on $t_0$. Since
\be
\Mpl^2 R = \frac{48 \Mpl^2}{c^4 t_0^2} \Bigl(1-\frac{c^2}{4} \Bigr) \, ,
\ee
requiring that the effective curvature remains below some cutoff ${\cal M}_{UV}^4$ imposes
\be
t_0^2 \ga \frac{48 \Mpl^2}{c^4 {\cal M}_{UV}^4} \Bigl(1-\frac{c^2}{4} \Bigr) \, .
\label{cutoff}
\ee
Since ${\cal M}_{UV} \la \Mpl/\sqrt{N}$ where $N$ is the number of light field theory species 
\cite{Jacobson:1994iw,Dvali:2007wp}, and $|c| \ll 1$, this implies
that $t_0 \gg 1/\Mpl$. We will see that this essentially pushes the selfreproduction regime of inflation too close to singularity,
and cuts it out of the semiclassical regime.  

Let us now turn to observables. The scalar and tensor perturbations spectra evaluated on the attractor are 
\be
P_{\tt S} = \Bigl(\frac{H^2}{2\pi \dot \phi}\Bigr)^2  \, , ~~~~~~~~~~~~~ P_{\tt T} = 
\frac{8H^2}{(2\pi)^2 \Mpl^2} \, .
\label{power}
\ee
Taking $t_*$ as the instant when the attractor evolution starts to 
dominate, corresponding to the 
value $\phi_*$, and introducing ${\cal N} = \ln(a(t)/a_*)$ as the number of efolds that transpired until time $t$, we find that the 
field variation is $\Delta \phi/M_{Pl} =  c {\cal N}$, and that the scalar power, tensor power, spectral index $n_S$ and the 
tensor-scalar ratio during this epoch are \cite{Mathiazhagan:1984vi,Lucchin:1984yf,Liddle:1988tb,Kalara:1990ar}
\be
P_{\tt S} = P_{\tt S}(t_*) \Bigl(\frac{k}{k_*}\Bigr)^{-2c^2/(2-c^2)} \, , ~~~~~~~ P_{\tt T} = r \, P_{\tt S} \, , 
~~~~~~~ n_S = 1- \frac{2c^2}{2-c^2} \, , ~~~~~~~ r = 8c^2 \, ,
\ee
where $P_{\tt S}(t_*), k_*$ are the Planck normalization values, $P_{\tt S}(t_*) \simeq 2.1 \times 10^{-9}$ \cite{planck18}. These formulas
are totally independent of ${\cal N}$, which is only determined by the variation of $\phi$ in the field space,
$c{\cal N} = \Delta \phi/M_{Pl}$. Note that as  consequence in these models 
the spectral running vanishes, $\alpha = \frac{dn_{\tt S}}{d\ln k} = 0$. These
examples are a special case of constant roll inflation \cite{Motohashi:2014ppa}. 
If we normalize the parameters by setting $n_{\tt S} \simeq 0.965$ 
for the CMB anisotropies, we find $c \simeq 0.185$ and
$r \simeq 0.274$. As it stands, this is in conflict with bounds on $r$ from BICEP/Keck \cite{BICEP:2021xfz}, 
calibrated to plain vanilla $\Lambda$CDM late universe. 
However, since the exponential potential by itself can't be the whole story, 
after all needing corrections to accommodate reheating 
at the very least \cite{Garriga:1998ri}, those deviations could 
fit \cite{BICEP:2021xfz}. Alternatively, if the resolution of the
$H_0$ tension forces a modification of $\Lambda$CDM, by for example 
inclusion of the EDE \cite{Poulin:2018cxd,Niedermann:2019olb,Niedermann:2020dwg}, the primordial scalar spectrum may need to be slightly
modified to compensate for the change in the evolution of fluctuations 
\cite{Ye:2021nej,Takahashi:2021bti,DAmico:2021fhz}. 

For example, if 
we pick $c$ such that $r \la 0.036$, to match the bounds of \cite{BICEP:2021xfz}, we find 
\be
c \la 0.067 \, , ~~~~~~~~~ 
n_{\tt S} \ga  0.995 \, .
\label{fitede}
\ee 
To fit the CMB we may need a 
slightly higher scalar spectral
index $n_{\tt S} \sim 0.98$ -- $0.995$ \cite{Ye:2021nej,Takahashi:2021bti,DAmico:2021fhz}. 
This means that nominally the exponential potentials 
satisfying (\ref{fitede}) might still be in the game. For those values of
$c$, the power controlling the attractor expansion rate is $2/c^2 \ga 444$. 

Further, as noted above, the total variation of
$\phi$ for ${\cal N}$ efolds is 
\be
\Delta \phi = c \Mpl {\cal N} \, ,
\ee
which for ${\cal N} \sim 60$ yields $\Delta \phi \simeq 4 \Mpl$, in some
tension with the purported swampland bounds 
\cite{Agrawal:2018own}, but not much. We will not worry too much about this issue here. 
We do note, however, that for these values of parameters, the bound on the cutoff $t_0$ of 
Eq. (\ref{cutoff}) leads to
\be
t_0 \ga \frac{4\sqrt{3}}{c^2} \Bigl(\frac{\Mpl}{{\cal M}_{UV}}\Bigr)^2 \Mpl^{-1} \, .
\label{cutoffbound}
\ee
At earlier times $t < t_0$, quantum gravity is doing most of the driving. 

\section{Causal Structure} 

We now turn to the causal structure of the power law inflation models, following \cite{Hellerman:2001yi,Fischler:2001yj}. 
Our particular interest is in the maximally extended past of the solutions with $k=0, -1$. We already know that the
geometries with power law scale factor are singular at $t\rightarrow 0$, but the question is, what kind of a singularity
is that. For simplicity, we start with $k=0$, and extend the scale factor
\be
a(t) =  \(\frac{t}{t_0}\)^{\frac{2}{3(1+w)}} \, ,
\label{flsol}
\ee
over the whole real semiaxis $(0,\infty)$. At future infinity, this scale factor is unbounded; however the curvature goes to zero and
locally the flat space approximation becomes ever better. To understand the global picture, we look at the Penrose diagram
describing such spacetimes. To obtain it, we  conformally map the solution 
on a section of the Einstein static universe, which is a direct product $R \times S^3$ 
with the metric 
\be
ds^2 = - d\tau^2 + d\chi^2 + \sin^2(\chi)d\Omega_2 \, . 
\ee
The section of $R \times S^3$ which describes power law inflation 
is the region bounded by the images of the singularities and/or past and
future causal boundaries. 

We find the required conformal map as a composition of two maps. 
First we transition to the conformally flat metric $ds^2 =
\omega^2(\bar x) \eta_{\mu\nu} d\bar x^\mu d\bar x^\nu$. In the second step, 
we map this metric to the static Einstein. The first map comprises of 
changing coordinates by 
\be
(1+3w)\frac{\bar t}{t_0} = 3(1+w)
\(\frac{t}{t_0}\)^{\frac{1+3w}{3(1+w)}} \, , ~~~~~~~~~~ \omega(\bar t) = 
\Bigl(\frac{1+3w}{3(1+w)} \frac{\bar t}{t_0} \Bigr)^{\frac{2}{1+3w}} \, .
\label{conffl}
\ee
When $-1 < w < -1/3$, the coordinate $\bar t$ 
is negative and inversely
proportional to $t$, varying from $-\infty$ to $0$ as
$t$ changes from $0$ to $\infty$: the $\bar t$-axis has the same
orientation as the $t$-axis.

The second map is defined by
\be 
\frac{r}{t_0} = \frac12 \Bigl(\tan(\frac{\chi+\tau}{2})
+ \tan(\frac{\chi-\tau}{2})
\Bigr) \, , ~~~~~~~~ \frac{\bar t}{t_0} = \frac12 \Bigl(\tan(\frac{\chi+\tau}{2}) -
\tan(\frac{\chi-\tau}{2}) \Bigr) \, .
\label{confes}
\ee
Since $r \in [0,\infty)$, $\bar t \in (-\infty,0)$, and
$\chi \in [0,\pi]$, it follows that $\tau \in
[-\pi,0]$. Putting together these formulas, the flat power law inflation metric is 
\be
ds^2 = {\cal C}^2 t_0^2 ~ \frac{[\cos(\frac{\chi-\tau}{2})
\cos(\frac{\chi+\tau}{2}) ]^{\frac{4}{|1+3w|}-2}}{~4~\sin^{\frac{4}{|1+3w|}}(|\tau|)}
\Bigl(-d\tau^2 +
d\chi^2 + \sin^2(\chi) d\Omega_2 \Bigr) \, , 
\label{esu}
\ee
where ${\cal C}$ is an ${\cal O}(1)$ constant, and
\be
\frac{|1+3w|}{6(1+w)}
\(\frac{t}{t_0}\)^{\frac{|1+3w|}{3(1+w)}} =
\Bigl(\tan(\frac{\chi-\tau}{2})
-\tan(\frac{\chi+\tau}{2}) \Bigr)^{-1} \, .
\label{ttau}
\ee

Using this, we see that the ultimate future of power law inflation, $t \rightarrow \infty$, for any fixed value of $r$, maps
onto $\tan(\frac{\chi-\tau}{2}) = \tan(\frac{\chi+\tau}{2})$: i.e precisely the latitude circle $\tau = 0$ on the cylinder.
Because the spacetime ends there, we cut out the portion of the cylinder $R \times S^2$ above it.
On the other hand, the singularity corresponds to the limit 
$t \rightarrow 0$ for any fixed $r$. By (\ref{confes}), (\ref{ttau}), we see that
it maps onto the curve $\tan(\frac{\chi-\tau}{2}) \rightarrow \infty$, which corresponds to 
$\tau = \chi - \pi$. It is clear that this is the null semi-circle connecting the
points $(-\pi,0)$ and $(0,\pi)$ on the cylinder. Since this hypersurface is
the ultimate singular past of power law inflation, we must throw out the portion of the
cylinder beneath it. Then we unwrap what remains, and find the causal
structure of Fig. ({\ref{fig1}).

\begin{figure}[thb]
    \centering
    \includegraphics[width=7.5cm]{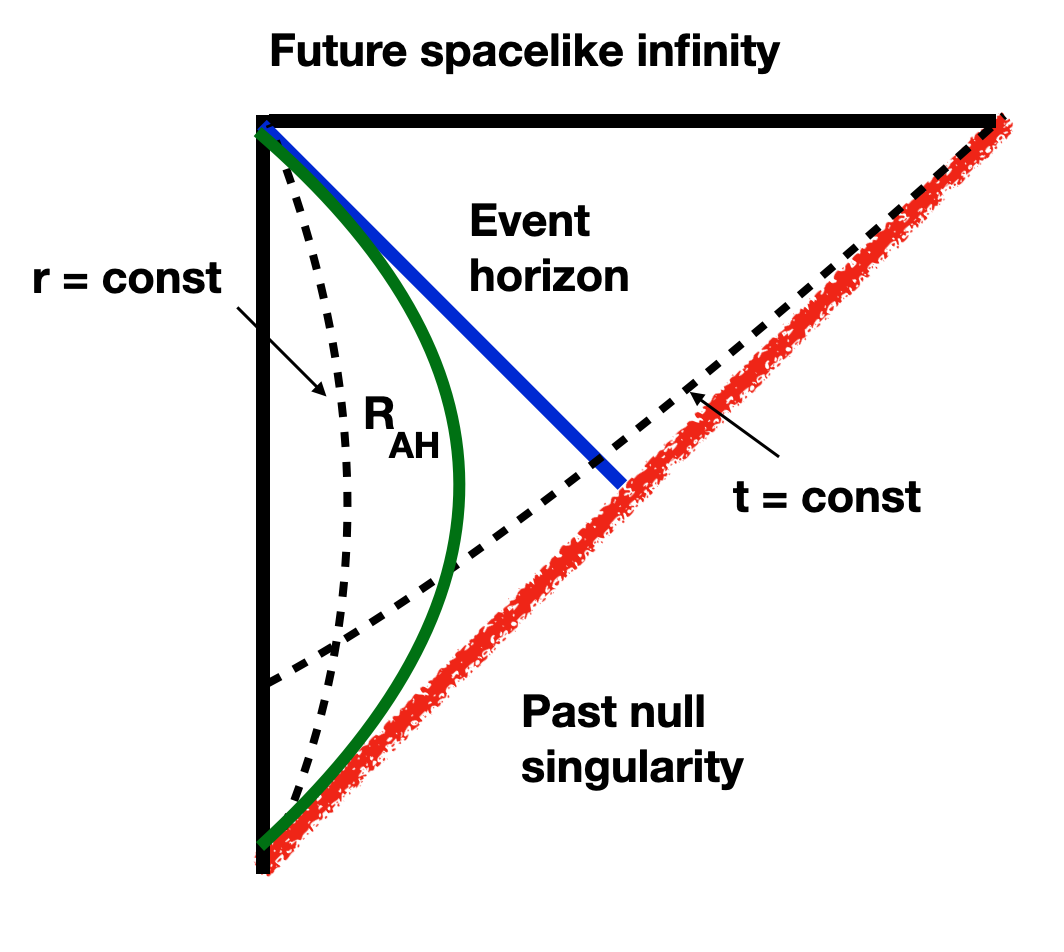}
    \caption{Causal structure of a spatially flat endless power law inflation. 
    Depicted are the event horizon, the apparent horizon ${\cal R}_{AH}$ and
    $r= {\rm const.}$ and $t = {\rm const.}$ hypersurfaces.}
    \label{fig1}
\end{figure}

Each point in Fig. (\ref{fig1}) corresponds to an angular $S^2$.
The ultimate past, which realizes the outcome of the Borde-Guth-Vilenkin theorem \cite{Borde:2001nh}, 
is a {\it null singularity}. If power law inflation never ends, the future is a spacelike
infinity. Any observer must have a future horizon, which in their rest frame is 
the null inward line ending in the upper left corner of
the diagram. Any observer would
find the universe at any given finite time
to be of finite size, being able to causally
explore only the interior of the diamond bounded by the horizon
and the singularity. 

The causal structure analysis so far concerns spatially flat geometry $k=0$. What if the universe is
open, $k=-1$? As we noted above, in this case, the expansion rate is set by a competition between 
spatial curvature and the exponential potential. At late times, the potential wins because of the attractor
behavior. However, early on the curvature can be dominant. When that happens, the scale
factor is a linear function of the comoving time, $a = t/t_0$. In this limit the metric is
\be
ds^2 = -dt^2 +  \(\frac{t}{t_0}\)^2 \Bigl(\frac{dr^2}{1+r^2} + r^2
d\Omega_2 \Bigr) \, .
\label{openbc}
\ee
At first glance one might think the metric is locally 
just a Milne wedge of the flat Minkowski in an accelerated
reference frame. However, thanks to $t_0$ this is not so: there is a 
real curvature singularity at $t \rightarrow 0$. The singularity is again
null, as we can see by mapping the slice of the spacetime near $t=0$ onto
the static Einstein universe. In this case the analogue of eq. (\ref{conffl}) is
\be 
\bar t = t_0  \ln(t/t_0) + \ldots \, , ~~~~~~~~~~ \omega(\bar t) =  e^{\frac{\bar t}{t_0}}  + \ldots \, .
\label{bconffl}
\ee
and so 
\be
\ln(t/t_0) + \ldots = \frac{1}{2}
\Bigl(\tan(\frac{\chi+\tau}{2}) -
\tan(\frac{\chi-\tau}{2}) \Bigr) \, .
\label{logt}
\ee
The ellipses denote the subleading terms when $t \rightarrow 0$. 
Hence the singularity again maps on the past null semi-circle $\tau = \chi -
\pi$.

At larger values of $t$ this geometry changes into the attractor-controlled section, 
where the curvature is locally negligible. 
If power law inflation lasts forever, the Penrose diagram is very similar to Fig (\ref{fig1}),
except for the local differences near the null singularity, as depicted in Fig (\ref{fig2}).
\begin{figure}[thb]
    \centering
    \includegraphics[width=6cm]{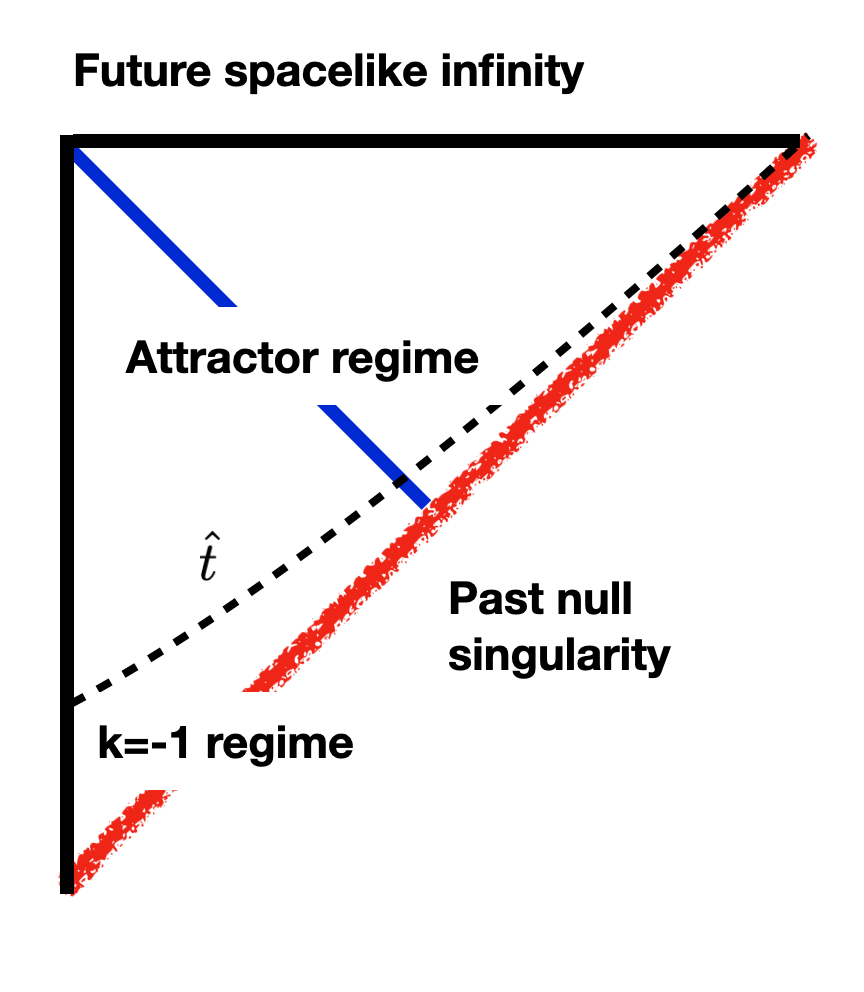}
    \caption{Causal structure of a spatially open power law inflation. It is an amalgam of the past $k=-1$ regime
    and a future power law attractor, matched together at a time $\sim \hat t$ (which in reality is a slab of worldvolume 
    few Hubble times thick).}
    \label{fig2}
\end{figure}

Clearly, both of these cases are reminiscent of the spatially flat charts of de Sitter,
with the exception that the past horizon is replaced by a null singularity. Nevertheless
as long as the metrics are purely FRW -- isotropic and homogeneous -- 
Weyl tensor vanishes there. Unlike in de Sitter the future horizon is not at 
constant spatial separation from the observer, but grows according to ($w<-1/3$)
\be
L_H = a(t) \int^{\infty}_t \frac{dt'}{a(t')}  = \frac{3(1+w)}{|1+3w|} t \, ,
\label{tdeph}
\ee
which shows that the volume of any spacelike
hypersurface inside the causal diamond grows extremely large. Yet the volume outside 
grows even larger \cite{Hellerman:2001yi,Fischler:2001yj}.

The cosmic inventory, as tallied by a single observer who receives the signals from their past, can be
accounted for by the capacity of the holographic screen, which is bounded by the area of the apparent 
horizon \cite{Fischler:1998st,Bousso:1999xy,Bousso:1999cb}. The apparent horizon ${\cal R}_{AH}$ is a boundary of 
the normal region of space, which colloquially we may think of the largest region that behaves as a 
locally Minkowski space. Specifically, it is the largest region inside which the beams of all outward geodesics, 
future or past oriented, spread out. On the apparent horizon, at least one class refocuses. This means,
the apparent horizon behaves like a lens. In our case, the exterior of the apparent horizon in all our
examples is an anti-trapped region, meaning that all past oriented null geodesics outside
of the apparent horizon, inward or outward bound, are converging. This is because of the null singularity
in the past. 

To find the location of the apparent horizon, recall that it is the hypersurface where at least one family
of null lines has vanishing expansion. If we consider a sphere of radius $ar$
with area $A \sim a^2(t) r^2$,  along the radial null geodesics
$dt = \pm a(t) dr$, the gradient of $A$ is $A' \sim a'r + ar'$
where the prime denotes the derivative with respect to the
affine parameter of the null line. The extremum yields the comoving size of the apparent horizon 
to be $r = 1/\dot a(t)$, and
so the proper apparent horizon size is\footnote{In truth, ${\cal R}_{AH} = 1/\sqrt{H^2 + k/a^2}$, but we
neglect the curvature term assuming the attractor to be a long stage. In the regime where the
curvature term dominates over the scalar the variation of ${\cal R}_{AH}$ is slower than linear, but
it still goes to zero on the singularity.}
\be
{\cal R}_{AH} = \frac1{H} = \frac{3(1+w)}{2} t \, .
\label{ah}
\ee
Clearly, since ${\cal R}_{AH}/L_H = |1+3w|/2 < 1$ for $-1<w < -1/3$, ${\cal R}_{AH}$
is always inside the future horizon.
On the diagram of Fig. (\ref{fig1}), it is the
arc ${\cal R}_{AH}$ between the lower left corner and the upper left
corner. 

Given the discussion above, it should be obvious that to get a realistic cosmology out of power law inflation, we 
need to end inflation and reheat the universe. We also need to perturb the reheating surface with the scalar and, unavoidably, tensor
fluctuations which we discussed in the previous section. Under those conditions, it is easy to see that the causal structure of
such a universe is represented by the Penrose diagram of Fig. (\ref{fig3}). There we allow for the possibility that very early on
the universe is open and curvature dominated, then transitions to the attractor regime, which ends globally with reheating. The reheating
surface will be smooth only down to one part in 10000, due to the inflationary fluctuations. As we discussed in the preceding section,
the dynamics can match the observations, with some tweaks.  
\begin{figure}[thb]
    \centering
    \includegraphics[width=6cm]{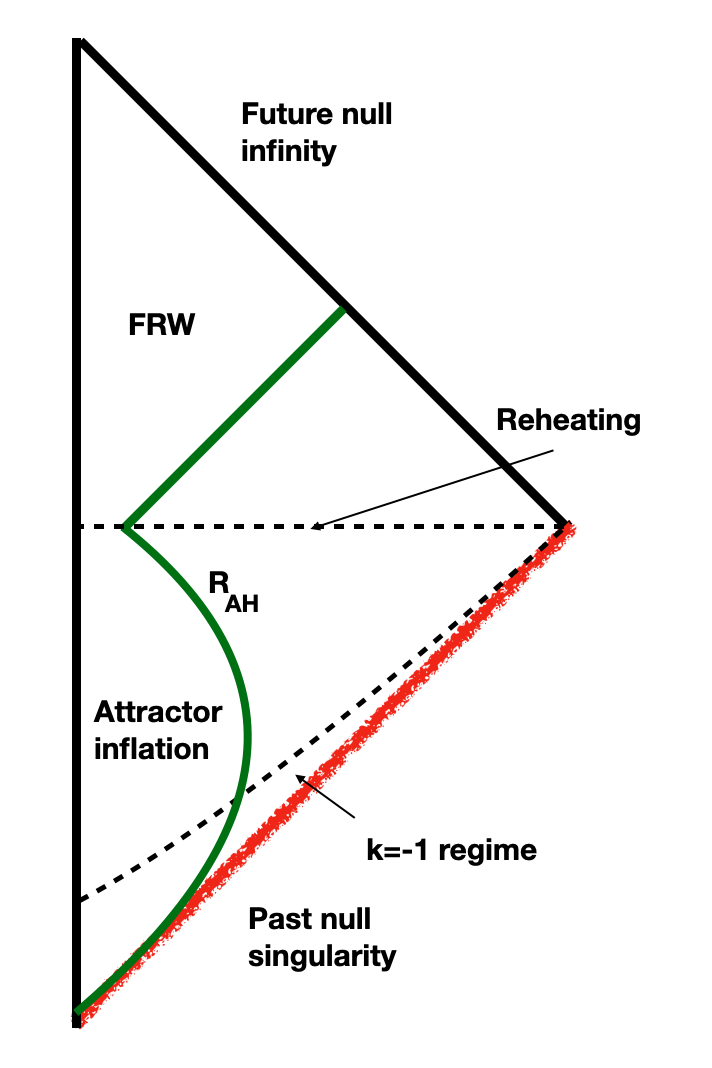}
    \caption{Causal structure of a spatially open power law inflation 
    which exits to radiation and matter dominated FRW. It is an amalgam of 
    the past $k=-1$ regime, a future power law attractor, and the postinflationary decelerating FRW. }
    \label{fig3}
\end{figure}

It is interesting to get an idea of the complexity of this universe at its various stages. We pick an observer and let them count what they
see, from the comfort of their rest frame. They do so by collecting the photons arriving from afar, and originating from as early as near the
null singularity (or gravitons instead of photons, since the universe is far more transparent for those). As those null rays approach
the observer -- they are future oriented inward geodesics -- they cross the apparent horizon and focus to the origin in the normal
region of spacetime surrounding the observer. The total amount of information coming in must satisfy the horizon area
bound \cite{Fischler:1998st,Bousso:1999xy,Bousso:1999cb}, $S \la {\cal A}_{AH}/4G_N$. Since the apparent horizon expands,
the information contents grows, but during the accelerated epoch the variation is very slow. 
The apparent horizon area evolves according to ${\dot {\cal A}_{AH}}/{{\cal A}_{AH}} = c^2 {H}$, which by using ${\cal N} = \ln(a/a_*)$ 
we can express as variation per efold,
\be
\frac{d {\cal A}_{AH}}{{\cal A}_{AH}} = c^2 d {\cal N} \, , 
\label{apphorarea}
\ee
and so during the attractor stage since $c \ll 1$, the maximal entropy is growing very slowly. 
This slow increase\footnote{Which could
be associated with the horizon crossing of the perturbations \cite{Albrecht:2002xs,Arkani-Hamed:2007ryv}.} continues until
the end of inflation, after which the growth rate changes to $d {\cal A}_{AH}/{\cal A}_{AH} = {\cal O}(1) d{\cal N}$, with
the precise details being controlled by the post-inflationary cosmic inventory.

The evolution in the semi-classical regime being adiabatic, with a globally fixed arrow of time as selected by the null singularity, 
means the ``entropy" is crossing the apparent horizon during inflation very slowly, 
and after inflation much more rapidly. Still, 
close to the singularity the geometry may still undergo a phase of 
selfreproduction. If so then different segments of the attractor regime of inflation could be subject to different perturbations, 
that can trigger the onset of exit at different times, or perhaps even prevent it altogether. If so those phenomena could alter the 
arrow of time in some parts of the spacetime. However, if we demand that the attractor dynamics yields observables
close to the current limits, the selfreproduction regime is excised out of the semiclassical limit. We can verify this as follows.
The boundary of selfreproduction is approximately given by the field values where $P_{\tt S} \simeq 1$, or more accurately 
the equality of the classical field variation integrated over a Hubble
time and the quantum fluctuation induced by cosmic acceleration,
\be
\int_{Hubble ~time} d\phi \simeq \frac{H}{2\pi} \, .
\ee
In other words, where the field variation is slow enough, the quantum Brownian drift can compensate it, and ``reboot" inflation. Using
$\dot \phi = \frac{2\Mpl}{ct}$ and $H = \frac{2}{c^2 t}$ yields for $c\ll 1$
\be
t_{\tt boundary} = \frac{1}{\pi c^3} \Mpl^{-1} \, 
\label{tbound}
\ee
Selfreproduction could only occur for $t < t_{\tt boundary}$, and the slow roll regime of inflation for $t  > t_{\tt boundary}$
(we could have phrased this condition in terms of the gauge invariant variable $\phi$ instead, but since
we gauge fixed the solution that is not necessary). However: our result for the cutoff $t_0$ of Eq. (\ref{cutoffbound}) severely obstructs
the selfreproduction regime. Namely, comparing (\ref{cutoffbound}) and (\ref{tbound}), 
\be
\frac{t_0}{t_{\tt boundary}} \ga  4\sqrt{3} \pi c \Bigl(\frac{\Mpl}{{\cal M}_{UV}}\Bigr)^2 \simeq 4\sqrt{3} \pi c N \, , 
\label{ratiobound}
\ee
where as we noted above $N$ is the number of light species in the theory, below the cutoff. If we take those to only
count the Standard Model degrees of freedom, $N \sim 120$, and so the right hand side is $\sim 2612 c$. If we further
require that $n_{\tt S}$ is not greater than $0.998$, we find $c \ga $0.044. This means that for the values of $c$ closest to
fitting the data, the ratio of Eq. (\ref{ratiobound}) is much greater 
than unity, $\frac{t_0}{t_{\tt boundary}} \gg  1$. Since only the time interval
$t > t_0$ is allowed in the effective theory, it means that the regime of selfreproduction is basically confined to the spacetime
sliver right next to the null singularity in Fig. (\ref{fig3}) that it is pointless to think about it. In other words, the selfreproduction regime
is behind the  Planckian cutoff surface above the null singularity, and it makes no sense physically in the solutions
depicted by Fig. (\ref{fig3}). As a result, the arrow of time, once set, remains preserved in those solutions. Taking
the solution to start from the null singularity as a homogeneous and isotropic FRW implies the vanishing of its Weyl tensor
in the far past. This will be violated later, by evolution, since quantum fluctuations of the scalar will perturb the geometry,
and this will contribute to the entropy production in the late universe. This is all fully consistent with Penrose's Weyl curvature
hypothesis. The question is, what selects this initial condition. 

\section{Cosmic Bubbles}

A rationale for selecting the initial condition which approximates really well the null singularity
with vanishing Weyl tensor could be provided using the framework of no boundary proposal for quantum cosmology
\cite{Hartle:1983ai} and
weighing the probabilities by the tunneling wavefunction prescription 
for the initial conditions \cite{Vilenkin:1982de,Linde:1983mx}. 
We will argue below that the process which mediates the creation of the universe depicted by the causal
structure of Fig. (\ref{fig3}) is closely related to the Hawking-Turok instanton \cite{Hawking:1998bn,Turok:1998he,Linde:1998gs,Unruh:1998wc,Vilenkin:1998pp}. 
We start by first briefly reviewing the Hawking-Turok instanton.

The idea is to imagine a theory of open inflating universe which tunnels from nothing, with a generic potential that can support 60 efolds of inflation.
This universe originates by a formation of a bubble of spacetime, and the universe is its dynamical interior. The pre-genesis stage is described by
an Euclidean geometry which resembles a squashed sphere \cite{Hawking:1998bn}. 
The scalar gradients will get large in some region of the Euclidean space, and produce a singularity which lies on the hypersurface of vanishing extrinsic curvature along 
which the analytical continuation is carried out \cite{Hawking:1998bn,Turok:1998he,Linde:1998gs,Unruh:1998wc,Vilenkin:1998pp}. In this regime,
the simple limit of relevant equations is 
\be
ds^2 = d\sigma^2 + b^2(\sigma) \Bigl(d\psi^2+ \sin^2 \psi d\Omega_2\Bigr) \, , 
\label{eumet}
\ee
for the Euclidean metric and
\be
\phi'' + 3\frac{b'}{b} \phi' = \partial_\phi V  \, , ~~~~~~~ \frac{b''}{b} = -   \frac{1}{3 \Mpl^2} \Bigl(\frac{\phi'^2}{2} + V\Bigr)  \, , ~~~~~~~  \frac{b'^2}{b^2} = \frac{1}{b^2} + \frac{1}{3 \Mpl^2} \Bigl(\frac{\phi'^2}{2} - V\Bigr)  \, ,
\ee
for the scalar and gravitational equations. The prime is a derivative with respect to $\sigma$. 
In this regime, the field $\phi$ is rolling in the upside-down potential $-V$. Let us initially consider a point where the
geometry is regular, and hence sufficiently close to it must be locally $R^4$. If we place the coordinate origin
at that point, near it we must have $b \rightarrow \sigma + \ldots$, and by symmetry $\phi' \rightarrow 0, \phi \rightarrow {\rm const}.$ (otherwise we would encounter
a singularity in $\phi''$, and consequently in $\phi$ too). Moving away from this point, $b$ grows, but at a rate which is decreasing due to the $b''$ equation. So $b$ reaches a maximum, and turns around.
Past it, the scalar derivatives grow fast for generic potentials, and take over, forcing $\phi$ to diverge at some $\sigma = \sigma_*$. In this limit
$b \rightarrow \bigl(\frac32 \frac{{\cal C}^2}{\Mpl^2}\bigr)^{1/6} (\sigma_* - \sigma)^{1/3}$ and
$\phi  \rightarrow {\rm const}. - \sqrt{\frac23} \Mpl \ln(\sigma_* - \sigma)$. Note that this behavior generalizes the
spherical limit $b = \sin \sigma$ which describes $\phi = {\rm const}.$, with a constant potential. 

The metric (\ref{eumet}), with these properties of $b$, can now be analytically continued in two steps. First, 
changing the latitude coordinate $\psi$ to $\psi = \pi/2 + i \tau$ at the equatorial hypersphere gives
\be
ds^2 = d\sigma^2 + b^2(\sigma) \Bigl(-d\tau^2+ \cosh^2 \tau d\Omega_2\Bigr) \, , 
\label{ancont1}
\ee
which describes an anisotropic cosmology just ``north" of the equator  \cite{Hawking:1998bn}, which has a timelike singularity
at $\sigma = \sigma_*$. This geometry also has a horizon at $\sigma = 0$, where its Euclidean counterpart had a regular
point. We can analytically continue across $\sigma = 0$, therefore, by using $\tau = i\pi/2 + \chi$ and $\sigma = it$,
while defining $a(t)= - i b(it)$ \cite{Hawking:1998bn}. Since $b$ has no singular points along the imaginary axis, $a(t)$ is well defined. 
The metric in this latter region is
\be
ds^2 = -dt^2 + a^2(t) \Bigl(d\chi^2+ \sinh^2 \chi d\Omega_2\Bigr) \, , 
\label{ancont2}
\ee
i.e. precisely an open universe. This is precisely the same metric as our metric of Eq. (\ref{met}), with one exception: here,
$t \rightarrow 0$ is a regular null hypersurface, a horizon rather than a null singularity. The singularity is now resolved, and
hides in the past of the horizon, as depicted in Fig. (\ref{fig4}) (see \cite{Hawking:1998bn,Turok:1998he,Linde:1998gs,Unruh:1998wc,Vilenkin:1998pp}). 
Notice that since the metrics are analytic continuations of each other, and eg. (\ref{eumet}) is conformally flat, Weyl tensor remains zero everywhere. This shows that the replacement of the null past singularity by a timelike regulator
which asymptotes to null does not affect the interpretation of the solutions.

\begin{figure}[thb]
    \centering
    \includegraphics[width=8cm]{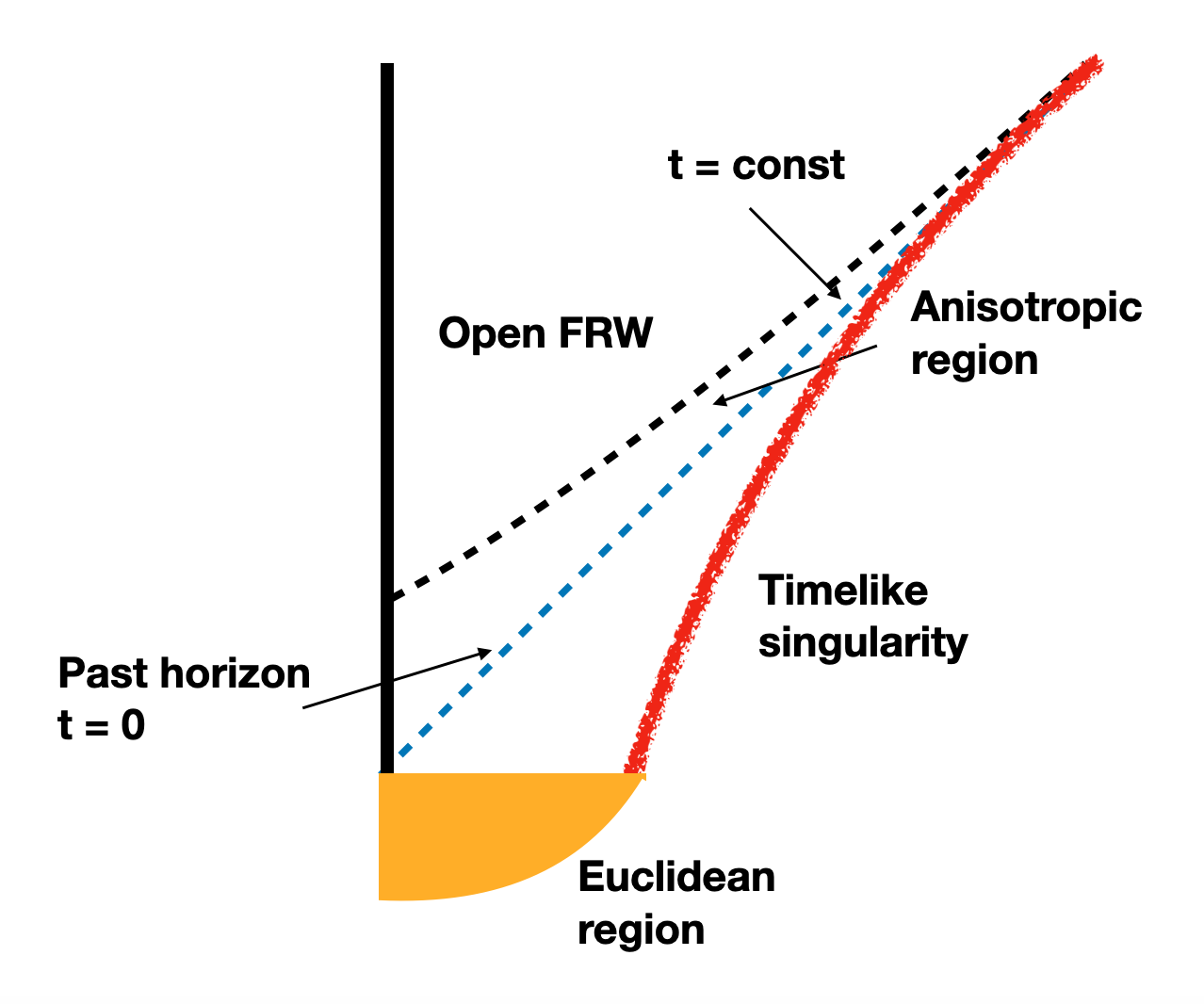}
    \caption{Resolving the singularity: on top, a spatially open power 
    law inflation which can exit to radiation and matter dominated FRW; $t=0$ null surface is now a 
    horizon. There is a timelike singularity behind it. This singularity can be excised by cutting out the region of
    space around it and replacing it with a bubble of flat space surrounded by a tensional domain wall \cite{Garriga:1998tm}, whose worldvolume lies between 
    the singularity and the $t=0$ horizon, or its generalizations \cite{Garriga:1998ri,Bousso:1998pk,Blanco-Pillado:2011fcm,Brown:2011gt}.
    }
    \label{fig4}
\end{figure}

In the final step, as in \cite{Garriga:1998tm,Garriga:1998ri,Bousso:1998pk,Blanco-Pillado:2011fcm,Brown:2011gt}
we excise the region around the singularity and replace it with a bubble, surrounded by a domain wall with some tension. 
We will not repeat all the technical procedure here, instead
referring the reader to the various options in \cite{Garriga:1998tm,Garriga:1998ri,Bousso:1998pk,Blanco-Pillado:2011fcm,Brown:2011gt}. 
The important point is that the worldvolume of the spherical domain wall asymptotes the past horizon from below. The surface energy 
density of the bubble is controlled by its initial size, and so the smaller it starts, the larger the density will be. In turn, this
controls the scale of the Euclidean action of the resolved configuration. 
For example in perhaps the simplest regularized case, where the bubble's interior is a ball of flat space, replacing the singular region.

Note that the metric surgery with cutting and pasting various pieces together across a domain wall
will not change the Weyl tensor of the configuration for the metrics which are $O(4)$ symmetric. The reason
is that the symmetry conditions are very restrictive for the metric, and only allow a single 
``free" function to appear in the metric - the scale factor, which is also the conformal factor.
It is the only term in the metric which picks up the boundary conditions. Thus the Weyl tensor remains
insensitive to the singularity regulator. If Weyl is zero without the regulator, it remains zero with it. 

Garriga found that the matching conditions $b'/b |_{\tt out} = - \kappa |{\cal C}|/3b^3$ and $b'/b |_{\tt in}  = 1/b$ with the 
bubble wall tension modeled by $\mu = \mu_0 - \alpha e^{\kappa \phi}$ yield the regulator contribution to the Euclidean action which is
\be
S_{\tt sing} = \frac13 S_{\tt GH} = \frac{\pi^2 |{\cal C}|}{\kappa} \, . 
\label{garry}
\ee
With the inflationary potential also included, one will find additional contributions. A very thorough survey of 
possible instantons and the actions which govern their nucleation rate is given in \cite{Bousso:1998pk}. In the case when the
Hawking-Turok instanton is regulated by a tensional domain, the full $O(4)$ Euclidean action is given by \cite{Bousso:1998pk}
\be
S_{HT/D} = - \frac{24\pi^2}{3\Mpl^2 H^2} \Bigl(1- \cos(H\sigma_m) \Bigr) \, .
\label{bchamb}
\ee
where $3 \Mpl^2 H^2 = U(\phi_{\tt initial})$ and $\sigma_m$ is the location of the domain wall which serves as a seam between two 
geometries. The trick used by \cite{Bousso:1998pk} to construct the regular solution is to orbifold around the wall\footnote{This is very similar to warped
braneworld constructions of, e.g. \cite{Lukas:1998yy,Randall:1999ee,Randall:1999vf,Kaloper:1999sm}.} instead to think of it
as a boundary between the Hawking-Turok solution in the bulk and a ball of flat space excising the singularity. This can be interpreted as a creation of
two jointed open universes, or by identifying the two, a single Hawking-Turok geometry with a singularity excised by a wall
at the end of the world. This actually may assist 
with obstructing the interpretation of the regulated solution as coming from a bubble of nothing in $6D$, which may be problematic for
its use as a regulator of Hawking-Turok processes \cite{Blanco-Pillado:2011fcm,Brown:2011gt}. We will not delve into this very interesting issue
any further here. Instead we will treat the action of (\ref{bchamb}) as an estimate of the Hawking-Turok nucleation rate, even thought
it is probably sensitive to the precise details of the UV completion of the configuration.

\begin{figure}[thb]
    \centering
    \includegraphics[width=6cm]{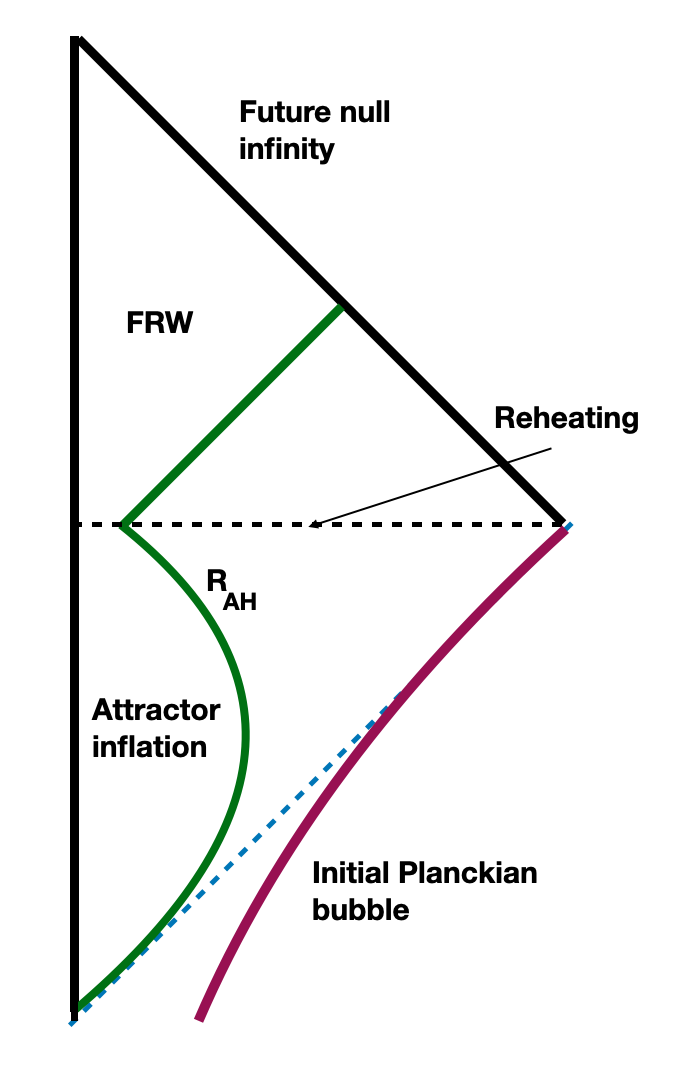}
    \caption{Resolving the singularity: the beginning a spatially open power 
    law inflation which exits to radiation and matter dominated FRW. The past 
    null singularity is interpreted as a domain wall of a bubble which initially nucleated at Planckian density. }
    \label{fig5}
\end{figure}

Note, that the wall will get closer -- ie approach its asymptotically null worldvolume -- the faster the smaller it starts, because it 
starts closer to the horizon initially. The regulated geometry is depicted in 
Fig. (\ref{fig5}). It should be clear that for all practical intents and purposes, 
if the universe arises as the interior of an initially small bubble, whose energy density is at or above the cutoff, this wall will behave 
practically as an almost null singularity: its worldvolume will be approximately null, and its energy density at the cutoff. Near the 
wall Weyl tensor will still be zero, and the geometry will behave to leading order just 
like the solution depicted in Fig. (\ref{fig3}). 

Using the tunneling from nothing probability prescription to estimate the likelihood of such a universe,  \cite{Vilenkin:1982de,Linde:1983mx}, 
\be
P \sim e^{S_{\tt euclidean}} \simeq  e^{-\frac{24\pi^2}{3\Mpl^2 H^2} \bigl(1- \cos(H\sigma_m) \bigr)} \, , 
\label{prob}
\ee
explains the selection of the initial conditions. First off, the $O(4)$ symmetry is favored over more complicated initial configurations 
by minimizing the action. Second, the initial values of $\phi$ which maximize the initial value of the potential are preferred over
those which make it small. Both of these conditions select inflationary history, and the exponential potential cuts off the possible
attainable number of efolds -- by not plateauing in the UV. This explains, at least in this context, how inflation starts\footnote{And addresses the issue of footnote 1.}.
Likewise, these conditions are also compatible with Penrose's conjecture,
since the initially $O(4)$ invariant geometry gives a vanishing Weyl tensor, and the (almost) null (regulated) singularity picks the
global time direction. One can then study entropy production, initially by studying metric perturbations, as in e.g. \cite{Garriga:1997wz,Garriga:1998he},
and later with the contributions from reheating and postinflationary evolution. The evolution of the apparent horizon area, Eq. (\ref{apphorarea}),
is consistent with post-nucleation entropy growth. Since many of the specific details can be found in the literature, we will not delve into the
details here.

\section{Summary} 

In this article, we have presented an argument that Penrose's vanishing Weyl curvature hypothesis, along with the initial 
singularity in the universe, motivated by the entropy considerations and the observed global arrow of time, is actually 
consistent with the inflationary paradigm. As an example, we used power law inflation which initially starts with a Hawking-Turok
nucleation process, with likelihood described by the tunneling from nothing probability. Note that here we demonstrated the 
compatibility of Penrose's Weyl curvature hypothesis and inflation -- where by inflation we mean the (semi)classical 
evolution of the background {\it augmented} with the selection of tunneling from nothing probability as a theory of initial
conditions -- without explicitly showing a more microscopic origin of either of these premises. That suffices for our purposes here.
Going beyond this goal requires a more precise exploration of the realms of quantum gravity, not easily accessible by 
present means.

Curiously, the resulting dynamics could even be
in marginal agreement with the current data. We note however that similar conclusions should hold for other models of inflation 
which start with the universe in a small bubble. The presence of the past (null) singularity will be generic for flat or open FRW
universes in the extreme past whenever the field value and the potential in that regime are not exactly constant. The
gradients near the past horizon will induce a large backreaction, and require regularization. Thus the general conclusions presented here
may hold even for potentials which fit the data better.

\vskip.5cm

{\bf Acknowledgments}: 
We would like to thank A. Albrecht, A. Lawrence, M. Sloth and A. Westphal  
for useful comments and discussions. We would also like to thank
MITP, Mainz, Germany, for kind hospitality during the course of this work. 
We thank the anonymous referee for interesting questions. 
NK is supported in part by the DOE Grant DE-SC0009999.


\begin{thebibliography}{99}

\bibitem{Penrose:1979azm}
R.~Penrose,
``Singularities and time-asymmetry,'', in {\it General Relativity: An Einstein Centenary Survey},
edited by S. Hawking and W. Israel, Cambridge University Press, Cambridge UK 1979.
 
\bibitem{Penrose:1988mg}
R.~Penrose,
Annals N. Y. Acad. Sci. \textbf{571}, 249-264 (1989).

\bibitem{Davies:1983nf}
P.~C.~W.~Davies,
Nature \textbf{301}, 398-400 (1983). 

\bibitem{Page:1983uh}
D.~N.~Page,
Nature \textbf{304}, 39-41 (1983). 

\bibitem{Davies:1984qc}
P.~C.~W.~Davies,
Nature \textbf{312}, 524-527 (1984).

\bibitem{Albrecht:2002uz}
A.~Albrecht,
[arXiv:astro-ph/0210527 [astro-ph]].

\bibitem{Albrecht:2004ke}
A.~Albrecht and L.~Sorbo,
Phys. Rev. D \textbf{70}, 063528 (2004)
[arXiv:hep-th/0405270 [hep-th]].

\bibitem{Carroll:2004pn}
S.~M.~Carroll and J.~Chen,
[arXiv:hep-th/0410270 [hep-th]].

\bibitem{Carroll:2005it}
S.~M.~Carroll and J.~Chen,
Gen. Rel. Grav. \textbf{37}, 1671-1674 (2005)
[arXiv:gr-qc/0505037 [gr-qc]].

\bibitem{Gibbons:2006pa}
G.~W.~Gibbons and N.~Turok,
Phys. Rev. D \textbf{77}, 063516 (2008)
[arXiv:hep-th/0609095 [hep-th]].

\bibitem{Guth:1980zm}
A.~H.~Guth,
Adv. Ser. Astrophys. Cosmol. \textbf{3}, 139-148 (1987).

\bibitem{Linde:1981mu}
A.~D.~Linde,
Adv. Ser. Astrophys. Cosmol. \textbf{3}, 149-153 (1987).

\bibitem{Albrecht:1982wi}
A.~Albrecht and P.~J.~Steinhardt,
Adv. Ser. Astrophys. Cosmol. \textbf{3}, 158-161 (1987).

\bibitem{Borde:2001nh}
A.~Borde, A.~H.~Guth and A.~Vilenkin,
Phys. Rev. Lett. \textbf{90}, 151301 (2003)
[arXiv:gr-qc/0110012 [gr-qc]].

\bibitem{Lesnefsky:2022fen}
J.~E.~Lesnefsky, D.~A.~Easson and P.~C.~W.~Davies,
[arXiv:2207.00955 [gr-qc]].

\bibitem{Wald:1983ky}
R.~M.~Wald,
Phys. Rev. D \textbf{28}, 2118-2120 (1983). 

\bibitem{bartip1} 
J.~D.~Barrow and F.~J.~Tipler,
MNRAS 216, 395-402 (1985).

\bibitem{bartip2}
J.~D.~Barrow, G.~J.~Galloway and F.~J.~Tipler,  
MNRAS 223, 835-844 (1986).

\bibitem{Kaloper:2002uj}
N.~Kaloper, M.~Kleban, A.~E.~Lawrence and S.~Shenker,
Phys. Rev. D \textbf{66}, 123510 (2002)
[arXiv:hep-th/0201158 [hep-th]].

\bibitem{Kaloper:2002cs}
N.~Kaloper, M.~Kleban, A.~Lawrence, S.~Shenker and L.~Susskind,
JHEP \textbf{11}, 037 (2002)
[arXiv:hep-th/0209231 [hep-th]].

\bibitem{East:2015ggf}
W.~E.~East, M.~Kleban, A.~Linde and L.~Senatore,
JCAP \textbf{09}, 010 (2016)
[arXiv:1511.05143 [hep-th]].

\bibitem{Kleban:2016sqm}
M.~Kleban and L.~Senatore,
JCAP \textbf{10}, 022 (2016)
[arXiv:1602.03520 [hep-th]].

\bibitem{Clough:2016ymm}
K.~Clough, E.~A.~Lim, B.~S.~DiNunno, W.~Fischler, R.~Flauger and S.~Paban,
JCAP \textbf{09}, 025 (2017)
[arXiv:1608.04408 [hep-th]].

\bibitem{Kaloper:2018zgi}
N.~Kaloper and J.~Scargill,
Phys. Rev. D \textbf{99}, no.10, 103514 (2019)
[arXiv:1802.09554 [hep-th]].

\bibitem{Burgess:2020nec}
C.~P.~Burgess, S.~P.~de Alwis and F.~Quevedo,
JCAP \textbf{05}, 037 (2021)
[arXiv:2011.03069 [hep-th]].

\bibitem{Mathiazhagan:1984vi}
C.~Mathiazhagan and V.~B.~Johri,
Class. Quant. Grav. \textbf{1}, L29-L32 (1984). 

\bibitem{Lucchin:1984yf}
F.~Lucchin and S.~Matarrese,
Phys. Rev. D \textbf{32}, 1316 (1985). 

\bibitem{Liddle:1988tb}
A.~R.~Liddle,
Phys. Lett. B \textbf{220}, 502-508 (1989). 

\bibitem{Kalara:1990ar}
S.~Kalara, N.~Kaloper and K.~A.~Olive,
Nucl. Phys. B \textbf{341}, 252-272 (1990). 

\bibitem{Hellerman:2001yi}
S.~Hellerman, N.~Kaloper and L.~Susskind,
JHEP \textbf{06}, 003 (2001) 
[arXiv:hep-th/0104180 [hep-th]].

\bibitem{Fischler:2001yj}
W.~Fischler, A.~Kashani-Poor, R.~McNees and S.~Paban,
JHEP \textbf{07}, 003 (2001) 
[arXiv:hep-th/0104181 [hep-th]].

\bibitem{Hawking:1998bn}
S.~W.~Hawking and N.~Turok,
Phys. Lett. B \textbf{425}, 25-32 (1998)
[arXiv:hep-th/9802030 [hep-th]].

\bibitem{Turok:1998he}
N.~Turok and S.~W.~Hawking,
Phys. Lett. B \textbf{432}, 271-278 (1998)
[arXiv:hep-th/9803156 [hep-th]].

\bibitem{Linde:1998gs}
A.~D.~Linde,
Phys. Rev. D \textbf{58}, 083514 (1998)
[arXiv:gr-qc/9802038 [gr-qc]].

\bibitem{Unruh:1998wc}
W.~G.~Unruh,
[arXiv:gr-qc/9803050 [gr-qc]].

\bibitem{Vilenkin:1998pp}
A.~Vilenkin,
Phys. Rev. D \textbf{57}, 7069-7070 (1998)
[arXiv:hep-th/9803084 [hep-th]].

\bibitem{Aguirre:2003ck}
A.~Aguirre and S.~Gratton,
Phys. Rev. D \textbf{67}, 083515 (2003)
[arXiv:gr-qc/0301042 [gr-qc]].

\bibitem{Garriga:1998tm}
J.~Garriga,
Phys. Rev. D \textbf{61}, 047301 (2000)
[arXiv:hep-th/9803210 [hep-th]].

\bibitem{Garriga:1998ri}
J.~Garriga,
[arXiv:hep-th/9804106 [hep-th]].

\bibitem{Bousso:1998pk}
R.~Bousso and A.~Chamblin,
Phys. Rev. D \textbf{59}, 063504 (1999)
[arXiv:hep-th/9805167 [hep-th]].

\bibitem{Blanco-Pillado:2011fcm}
J.~J.~Blanco-Pillado, H.~S.~Ramadhan and B.~Shlaer,
JCAP \textbf{01}, 045 (2012)
[arXiv:1104.5229 [gr-qc]].

\bibitem{Brown:2011gt}
A.~R.~Brown and A.~Dahlen,
Phys. Rev. D \textbf{85}, 104026 (2012)
[arXiv:1111.0301 [hep-th]].


\bibitem{Coleman:1980aw}
S.~R.~Coleman and F.~De Luccia,
Phys. Rev. D \textbf{21}, 3305 (1980). 

\bibitem{Linde:2004nz}
A.~D.~Linde,
JCAP \textbf{10}, 004 (2004)
[arXiv:hep-th/0408164 [hep-th]].

\bibitem{BICEP:2021xfz}
P.~A.~R.~Ade \textit{et al.} [BICEP and Keck],
Phys. Rev. Lett. \textbf{127}, no.15, 151301 (2021)
[arXiv:2110.00483 [astro-ph.CO]].

\bibitem{Poulin:2018cxd}
V.~Poulin, T.~L.~Smith, T.~Karwal and M.~Kamionkowski,
Phys. Rev. Lett. \textbf{122}, no.22, 221301 (2019)
[arXiv:1811.04083 [astro-ph.CO]].

\bibitem{Niedermann:2019olb}
F.~Niedermann and M.~S.~Sloth,
Phys. Rev. D \textbf{103}, no.4, L041303 (2021)
[arXiv:1910.10739 [astro-ph.CO]].

\bibitem{Niedermann:2020dwg}
F.~Niedermann and M.~S.~Sloth,
Phys. Rev. D \textbf{102}, no.6, 063527 (2020)
[arXiv:2006.06686 [astro-ph.CO]].

\bibitem{Ye:2021nej}
G.~Ye, B.~Hu and Y.~S.~Piao,
Phys. Rev. D \textbf{104}, no.6, 063510 (2021)
[arXiv:2103.09729 [astro-ph.CO]].

\bibitem{Takahashi:2021bti}
F.~Takahashi and W.~Yin,
Phys. Lett. B \textbf{830}, 137143 (2022)
[arXiv:2112.06710 [astro-ph.CO]].

\bibitem{DAmico:2021fhz}
G.~D'Amico, N.~Kaloper and A.~Westphal,
Phys. Rev. D \textbf{105}, no.10, 103527 (2022)
[arXiv:2112.13861 [hep-th]].

\bibitem{Mukhanov:1996ak}
V.~F.~Mukhanov, L.~R.~W.~Abramo and R.~H.~Brandenberger,
Phys. Rev. Lett. \textbf{78}, 1624-1627 (1997)
[arXiv:gr-qc/9609026 [gr-qc]].

\bibitem{Polyakov:2007mm}
A.~M.~Polyakov,
Nucl. Phys. B \textbf{797}, 199-217 (2008)
[arXiv:0709.2899 [hep-th]].

\bibitem{Dvali:2017eba}
G.~Dvali, C.~Gomez and S.~Zell,
JCAP \textbf{06}, 028 (2017)
[arXiv:1701.08776 [hep-th]].

\bibitem{Kaloper:2022oqv}
N.~Kaloper,
Phys. Rev. D \textbf{106}, 065009 (2022)
[arXiv:2202.06977 [hep-th]].

\bibitem{Kaloper:2022utc}
N.~Kaloper,
Phys. Rev. D \textbf{106}, 044023 (2022)
[arXiv:2202.08860 [hep-th]].

\bibitem{Kaloper:2022jpv}
N.~Kaloper and A.~Westphal,
[arXiv:2204.13124 [hep-th]].


\bibitem{Jacobson:1994iw}
T.~Jacobson,
[arXiv:gr-qc/9404039 [gr-qc]].

\bibitem{Dvali:2007wp}
G.~Dvali and M.~Redi,
Phys. Rev. D \textbf{77}, 045027 (2008)
[arXiv:0710.4344 [hep-th]].

\bibitem{planck18}
Y.~Akrami \textit{et al.} [Planck],
Astron. Astrophys. \textbf{641}, A10 (2020) 
[arXiv:1807.06211 [astro-ph.CO]].

\bibitem{Motohashi:2014ppa}
H.~Motohashi, A.~A.~Starobinsky and J.~Yokoyama,
JCAP \textbf{09}, 018 (2015)
[arXiv:1411.5021 [astro-ph.CO]].

\bibitem{Agrawal:2018own}
P.~Agrawal, G.~Obied, P.~J.~Steinhardt and C.~Vafa,
Phys. Lett. B \textbf{784}, 271-276 (2018) 
[arXiv:1806.09718 [hep-th]].

\bibitem{Fischler:1998st}
W.~Fischler and L.~Susskind,
[arXiv:hep-th/9806039 [hep-th]].

\bibitem{Bousso:1999xy}
R.~Bousso,
JHEP \textbf{07}, 004 (1999)
[arXiv:hep-th/9905177 [hep-th]].

\bibitem{Bousso:1999cb}
R.~Bousso,
JHEP \textbf{06}, 028 (1999)
[arXiv:hep-th/9906022 [hep-th]].

\bibitem{Albrecht:2002xs}
A.~Albrecht, N.~Kaloper and Y.~S.~Song,
[arXiv:hep-th/0211221 [hep-th]].

\bibitem{Arkani-Hamed:2007ryv}
N.~Arkani-Hamed, S.~Dubovsky, A.~Nicolis, E.~Trincherini and G.~Villadoro,
JHEP \textbf{05}, 055 (2007)
[arXiv:0704.1814 [hep-th]].

\bibitem{Hartle:1983ai}
J.~B.~Hartle and S.~W.~Hawking,
Phys. Rev. D \textbf{28}, 2960-2975 (1983). 

\bibitem{Vilenkin:1982de}
A.~Vilenkin,
Phys. Lett. B \textbf{117}, 25-28 (1982). 

\bibitem{Linde:1983mx}
A.~D.~Linde,
Lett. Nuovo Cim. \textbf{39}, 401-405 (1984).

\bibitem{Lukas:1998yy}
A.~Lukas, B.~A.~Ovrut, K.~S.~Stelle and D.~Waldram,
Phys. Rev. D \textbf{59}, 086001 (1999)
[arXiv:hep-th/9803235 [hep-th]].

\bibitem{Randall:1999ee}
L.~Randall and R.~Sundrum,
Phys. Rev. Lett. \textbf{83}, 3370-3373 (1999)
[arXiv:hep-ph/9905221 [hep-ph]].

\bibitem{Randall:1999vf}
L.~Randall and R.~Sundrum,
Phys. Rev. Lett. \textbf{83}, 4690-4693 (1999)
[arXiv:hep-th/9906064 [hep-th]].

\bibitem{Kaloper:1999sm}
N.~Kaloper,
Phys. Rev. D \textbf{60}, 123506 (1999)
[arXiv:hep-th/9905210 [hep-th]].

\bibitem{Garriga:1997wz}
J.~Garriga, X.~Montes, M.~Sasaki and T.~Tanaka,
Nucl. Phys. B \textbf{513}, 343-374 (1998)
[erratum: Nucl. Phys. B \textbf{551}, 511-511 (1999)]
[arXiv:astro-ph/9706229 [astro-ph]].

\bibitem{Garriga:1998he}
J.~Garriga, X.~Montes, M.~Sasaki and T.~Tanaka,
Nucl. Phys. B \textbf{551}, 317-373 (1999)
[arXiv:astro-ph/9811257 [astro-ph]].

\end{thebibliography}
\end{document}